# Development of an LED reference light source for calibration of radiographic imaging detectors

M. Weierganz[1], D. Bar[2], B. Bromberger[1], V. Dangendorf [1], G. Feldman[2], M.B. Goldberg[1,2], M. Lindemann[1], I. Mor[2], K. Tittelmeier[1], D. Vartsky[2]

[1] Physikalisch-Technische Bundesanstalt, Braunschweig, Germany
[2] Soreq NRC, Yavne, Israel

## Abstract

A stable reference light source based on an LED (Light Emission Diode) is presented for stabilizing the conversion gain of the opto-electronic system of a gamma- and fast-neutron radiographic and tomographic imaging device. A constant fraction of the LED light is transported to the image plane of the camera and provides a stable reference exposure. This is used to normalize the images during off-line image processing. We have investigated parameters influencing the stability of LEDs and developed procedures and criteria to prepare and select LEDs suitable for delivering stable light outputs for several 100 h of operation.

## 1. Introduction

In recent years we have developed an imaging system for fast-neutron resonance radiography and tomography based on the „Time Resolved Integrative Optical Neutron“ (TRION) detector concept [1, 2]. The basic concept of TRION is the use of a fast scintillating fiber screen for neutron-to-optical-image conversion. The latter is viewed by fast, ns-gated image intensifiers. Each of these selects an image at one specific neutron energy by the time-of-flight method. Final image integration is via one or more individual CCD cameras. Details of the concept and performance results can be found in [1, 2]

For a quantitative analysis of the information contained in the resonance images – in particular for reconstructing the elemental distribution in the radiographed sample – it is of crucial importance that the sensitivity and electro-optical gain of the optical readout elements (image intensifiers and CCD cameras) are of the highest (few parts in $10^3$) stability. Alternatively, if this cannot be assured, one can monitor these parameters and normalize the radiographs in the subsequent image analysis procedure.

Image intensifiers are the most crucial electro-optical components in our system, serving as fast shutters and light amplification elements. Light amplification is obtained via electron multiplication in MCPs, whose gain is prone to smallest variations of the electric field across an MCP and, moreover, depends on the exposure history (due to channel – saturation, dead-time for field recovery inside a channel). Long term gain stability and rate-independent amplification can, in principle, be achieved with electronic stabilization techniques like e.g. the ones we have developed for photomultiplier tubes [3]. However, for the very fast changes in the radiation flux in the pulsed gamma/neutron beam we are using in our application such a technique is not applicable. Therefore, we have opted for an off-line correction of gain and sensitivity drifts, that apply not only to the intensifiers but to the entire optical chain.

To this end, a very stable LED light source was developed and mounted close to the edge of the fiber scintillator. This constant stabilized light source is viewed in parallel to the neutron image and provides each image with a stable reference for calibrating the effective sensitivity and gain of the full optical chain. The region of the screen beneath the reference light source (see figure 1) is optically shielded in order to reduce additional light contributions from the screen in this region. Moreover, stray light corrections are implemented in the algorithms that take into account contributions from reflected light inside the detector box, or for light and charge cross-talk in the intensifier(s).

The transport of the light (or, preferably, a small but stable fraction of it) from the stabilized LED source into the image plane of the optical chain also called for special measures to be taken. Initial tests with optical fibers proved unsuccessful because their light transmission depends significantly on temperature and on mechanical details such as bending radii of fibers or mechanical contacts to other materials touching the fiber (guard tubes, sockets, etc.). Therefore we finally used a simple, externally black anodized Al-tube of ~5 cm length with a radial bore of 1 mm diameter at the remote end to guide the light to a peripheral spot on the screen (see figure 1). The light intensity could be varied by inserting collimators into the tube. The size of the reference light spot can be adjusted to spot areas > 1 mm by defocusing, i.e. placing the source several  cm away from the screen (in the direction of the optical axis). Figure 1 shows a typical set up of the reference light source and the mechanical light guide mounted in front of the scintillating fiber neutron converter.

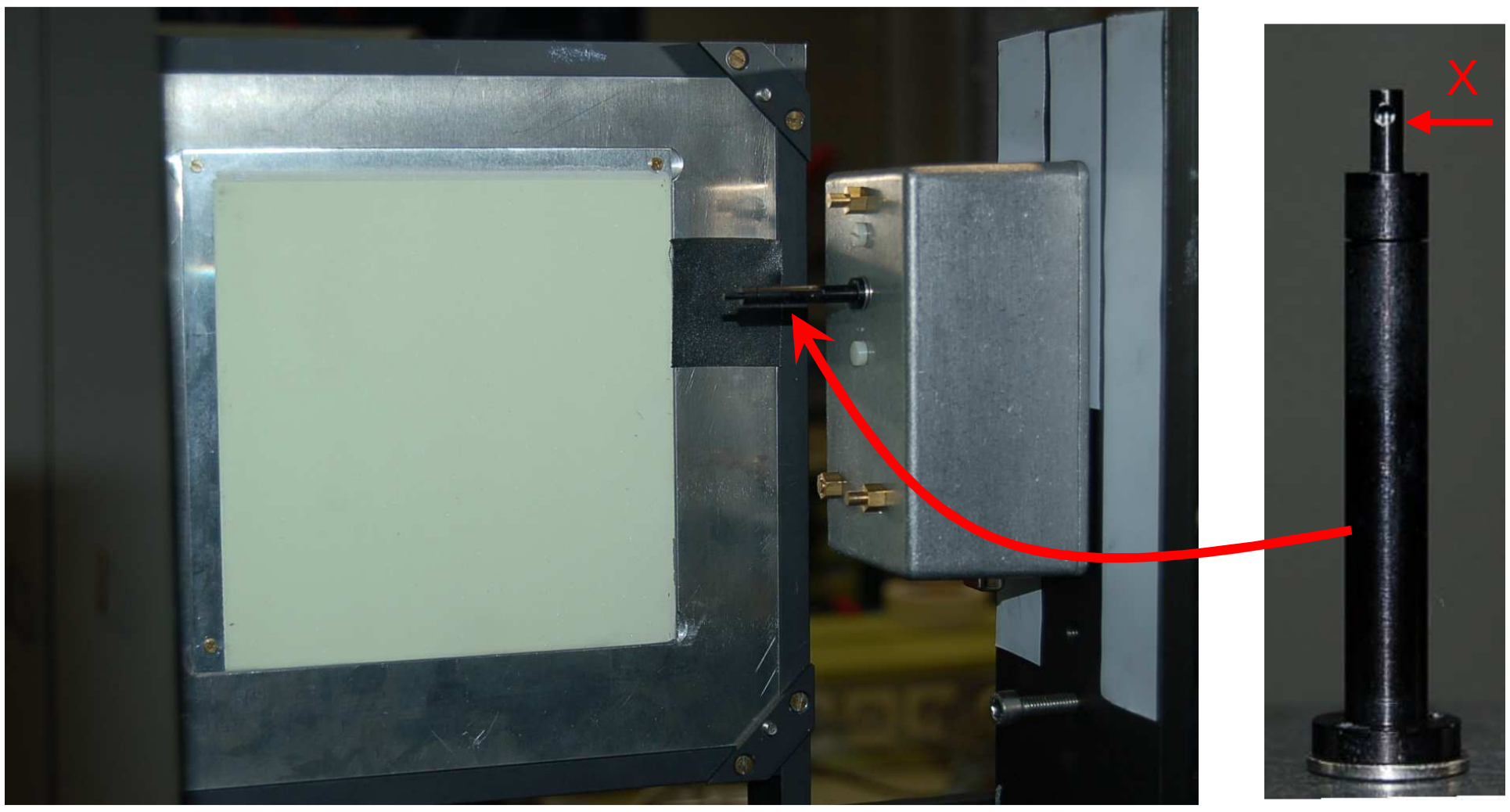

**Figure 1**. Photograph of the stabilized LED light source mounted in front of the scintillator (here a $^6$LiF/ZnS screen for thermal neutron radiography). Right side: enlarged photo of the light guide with the 1 mm opening for the light output at the side of the tip (marked "X")

A central task of the optics department at PTB is the derivation and dissemination of photometric quantities such as luminous intensity, luminous flux and the related colorimetric quantities: chromaticity coordinates, dominant wavelength, color temperature, etc. A central role in the dissemination or these quantities to users from science and industry is played by LEDs, mainly due to their comparatively high inherent stability and small size [4]. The LEDs distributed as transfer standard for the mentioned quantities are usually selected from standard commercial batches and aged for a few-hundred hours at a current close to their maximum current. After a careful, partly automated scrutinization, the most stable LEDs are selected and mounted into an electronic control- and stabilization device which guarantees stable  light output. The relative long term drift of their photometric properties is typically $< 10^{-4}\,h^{-1}$.

For the application in TRION these standards cannot be used directly because the electronics for control and stabilization is rather costly and bulky and cannot be integrated easily into the imaging system. Based on the experience of the optics group at PTB we therefore initiated a dedicated development, the goal being to design a compact and stable LED light source which runs self-sustained (i.e. without external electronics, diagnostics or computer) apart from a low voltage power supply, and can easily be integrated into the detector system

## 2. Design and implementation of the light source

### 2.1.Temperature-stabilization of the LED

One of the problems encountered in the use of LEDs as reference light-sources is the strong temperature-dependence of the light yield. Consequently, light-yield stability in the 0.1 %-range requires maintaining the temperature constant to within 0.1 ° limits. However, numerous manufacturer spec-sheets fail to quote the temperature-dependence characteristics altogether, or do so with insufficient precision. An example of the latter is the spec-sheet of Osram's LB543C [5], the informative value of which was inadequate for the purposes of the present application.

For example, measurements performed here with a white L5T20W [6] LED revealed a reduction of ~1% in light-output, when the temperature was raised from 45 ºC to 46 ºC. As shown in figure 2, the corresponding reduction measured for a blue LB543C was 0.5 %. It was accompanied by a loss of ~2.5 mV in diode forward-drop-voltage. This problem can be overcome to the required precision (in the 0.1 %-range) by providing an environment that is thermally-stabilized to within several 0.1 ºC.

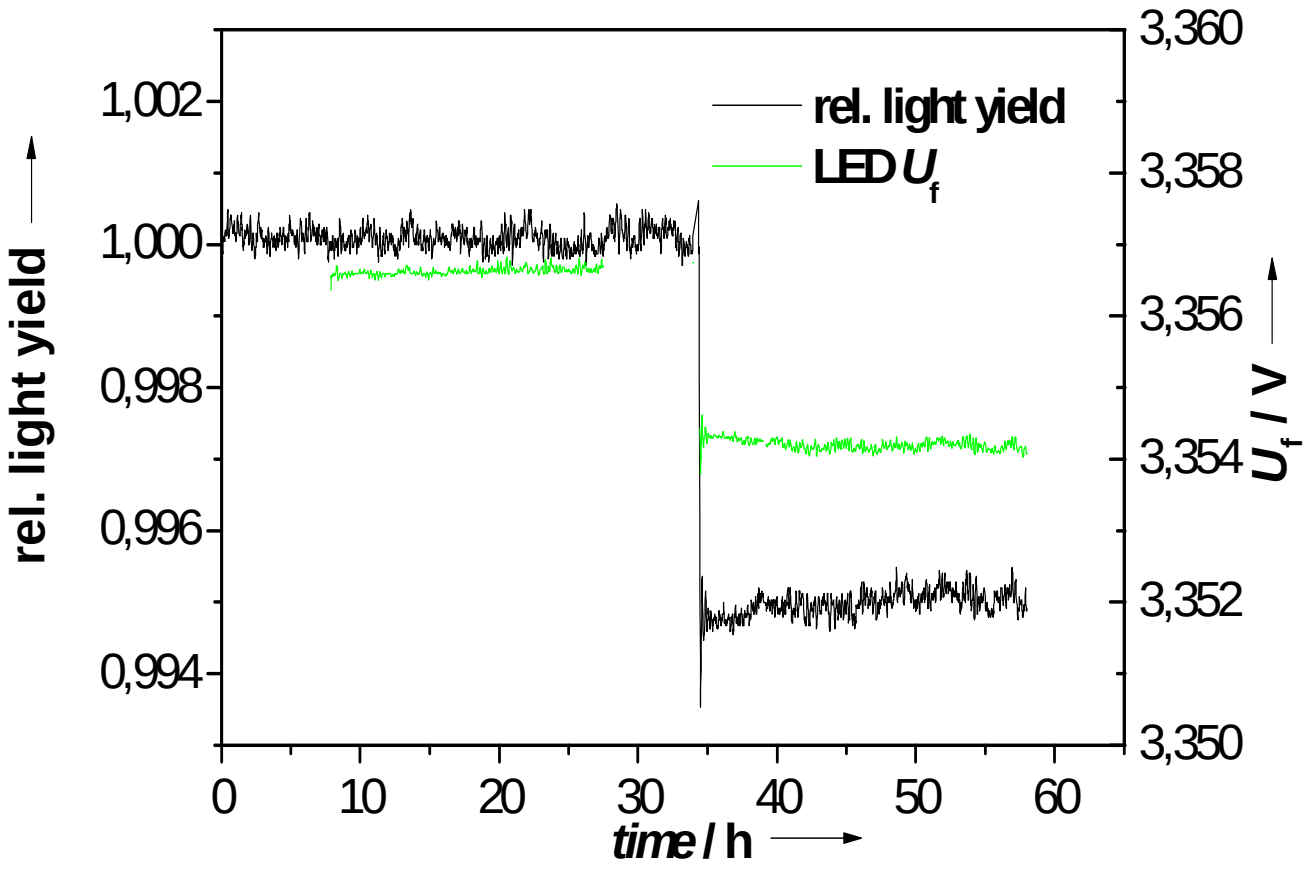

**Figure 2.** An increase in temperature from 45°to 46°C reduc es the light-yield of the LED LB543C by 0.5%.

The temperature control was implemented by mounting the LED on a transistor-heated copper block, schematically shown in figure 3. A polystyrene-foam insulator reduces the heat loss to the environment and thereby the thermal-gradient between temperature-sensor and LED. The temperature-sensor is cemented within a borehole inside the Cu-block. To reduce troublesome heat-losses, thin wire-coils were used for the electrical contacts to LED and temperature-sensor. A 1 mm bore was drilled through the block along an extension of the LED's main light-emission axis. A transverse 3 mm borehole permitted a monitor diode to monitor the LED light yield at 90º to the main optical axis. The entire setup, including temperature-controller and stabilized current-source, is shown in figure 4.

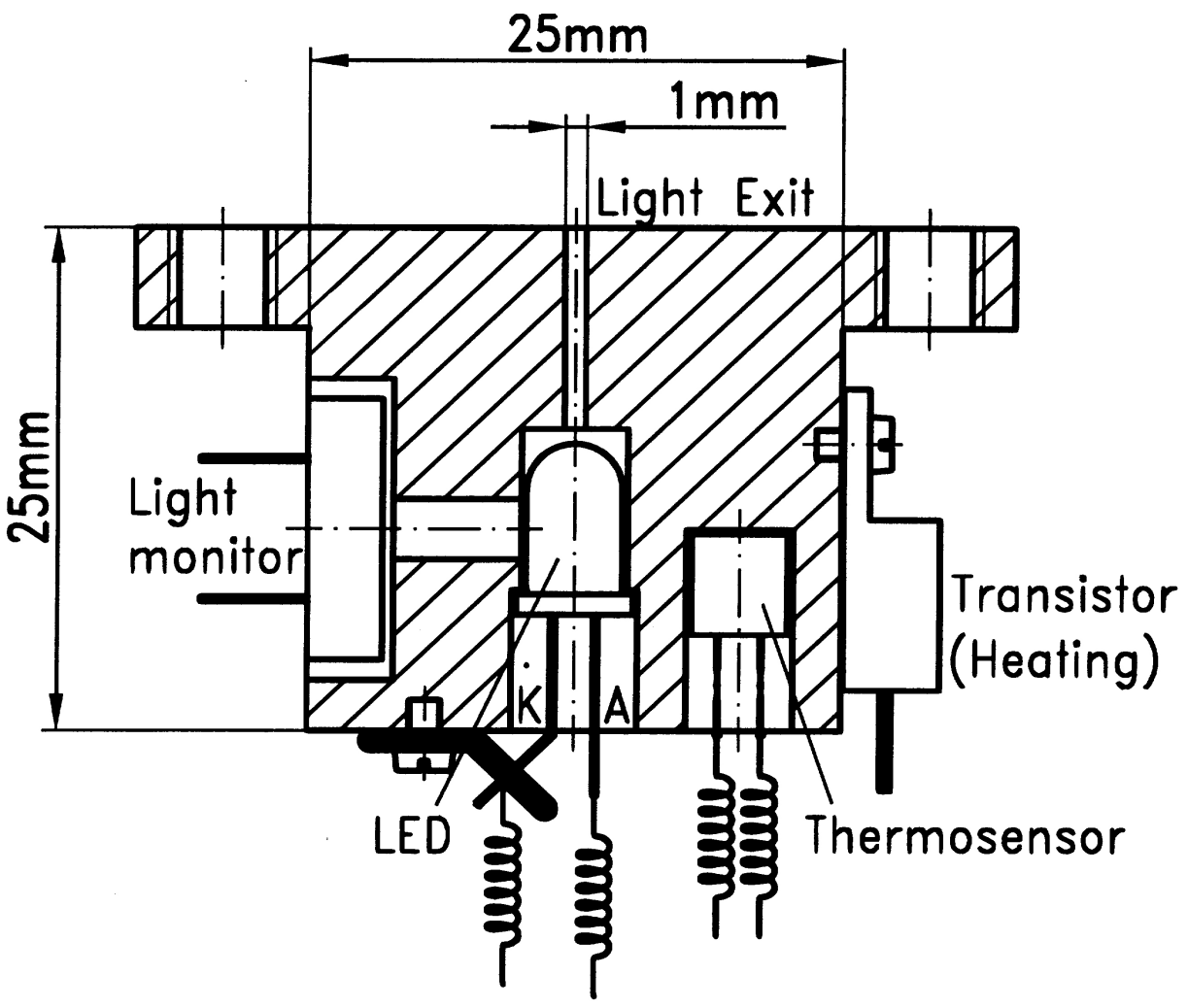

**Figure 3.** Cross-sectional view of the temperature stabilized copper block and the components attached to it.

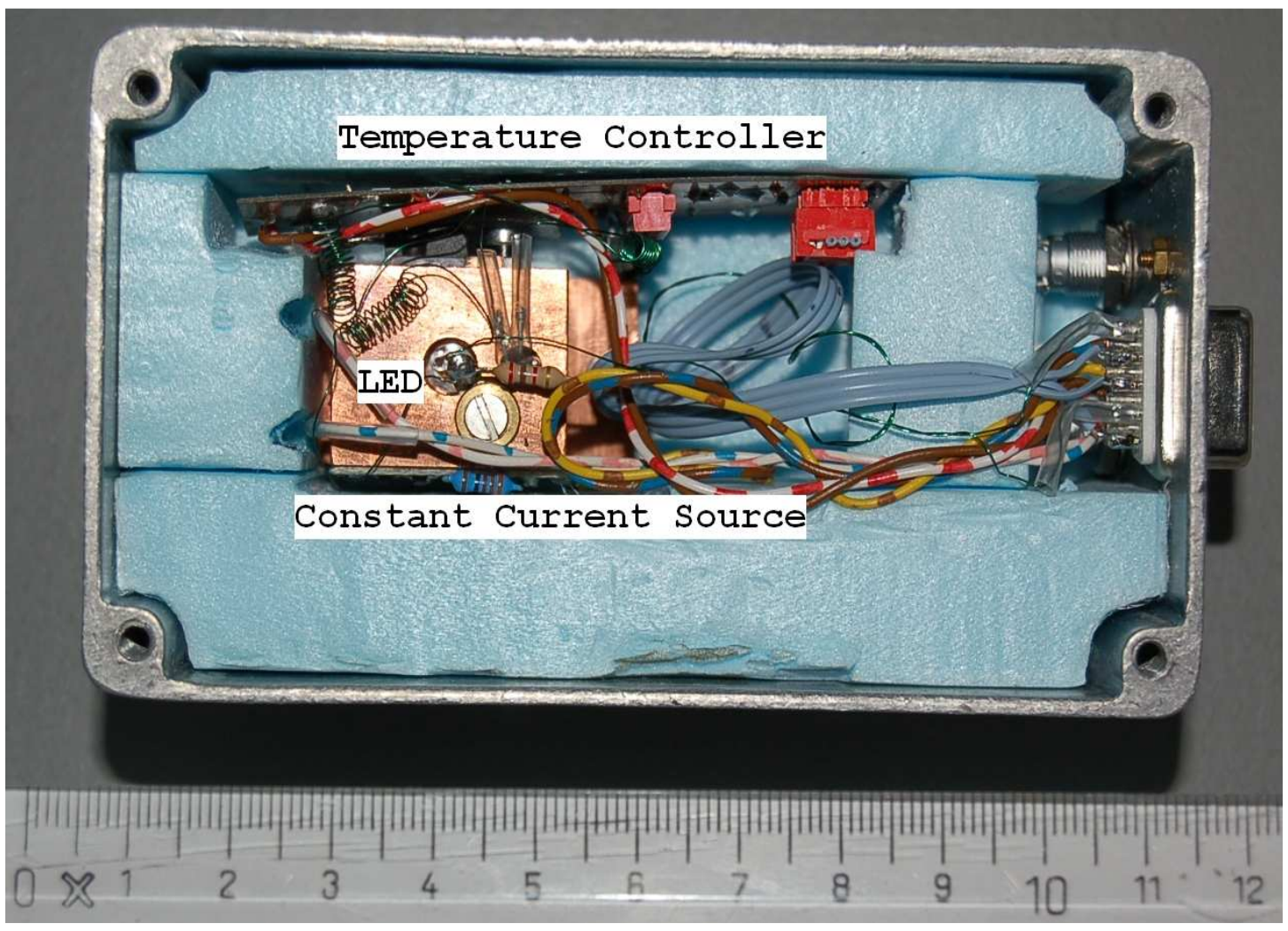

**Figure 4.** View of the reference light-source setup, including temperature-controller and stabilized current source.

Indeed, it is also possible to use the LED itself as temperature sensor, since the forward-drop-voltage is a function of the temperature. However, the value of this voltage changes also in the course of an (irreversible) aging process undergone by the LED during operation for extended periods (tens-to-hundreds of hours). Thus, it was decided to use the LM335 [7] external temperature-sensor instead.

A block diagram of the electronic circuit is shown in figure 5. A PI temperature-regulator was designed around the Atmel AT Mega8 Microcontroller [8], using a current-regulated TIP131 transistor [9] in a TO220-housing as heating element. The heating power is up to 3W, which ensured that the set-point temperature (typically 45 °C) was rapidly reached (see figure 6). Communication with this module (PI parameter values for the control circuit and the set-point temperature as well as monitoring of the actual temperature) is possible through an RS232-interface. However, it should be mentioned that after setting the parameters they are stored in the EEPROM of the processor and the unit operates independently of external control with the last-stored parameter set.

To prevent transients following a brief power-cut (the status of the integral-component in the micro-controller would be undefined after a restart), the updated integral-component values are periodically stored in the processor's EEPROM. Following a reset of the controller, the last-stored value of the integral (prior to the power-cut) would be adopted as the new initial value. With this procedure, power-cuts of a few seconds do not lead to appreciable transients in the control unit.

The effect of fluctuations of the ambient temperature on the output of the light-source is barely visible in figure 7. Specifically, an ambient temperature rise of 7 °C leads to only 0.15 % change in light-yield, i.e. 0.02 % / °C, compared to about 0.5 % / °C for a non-stabilized LED (see figure 2).

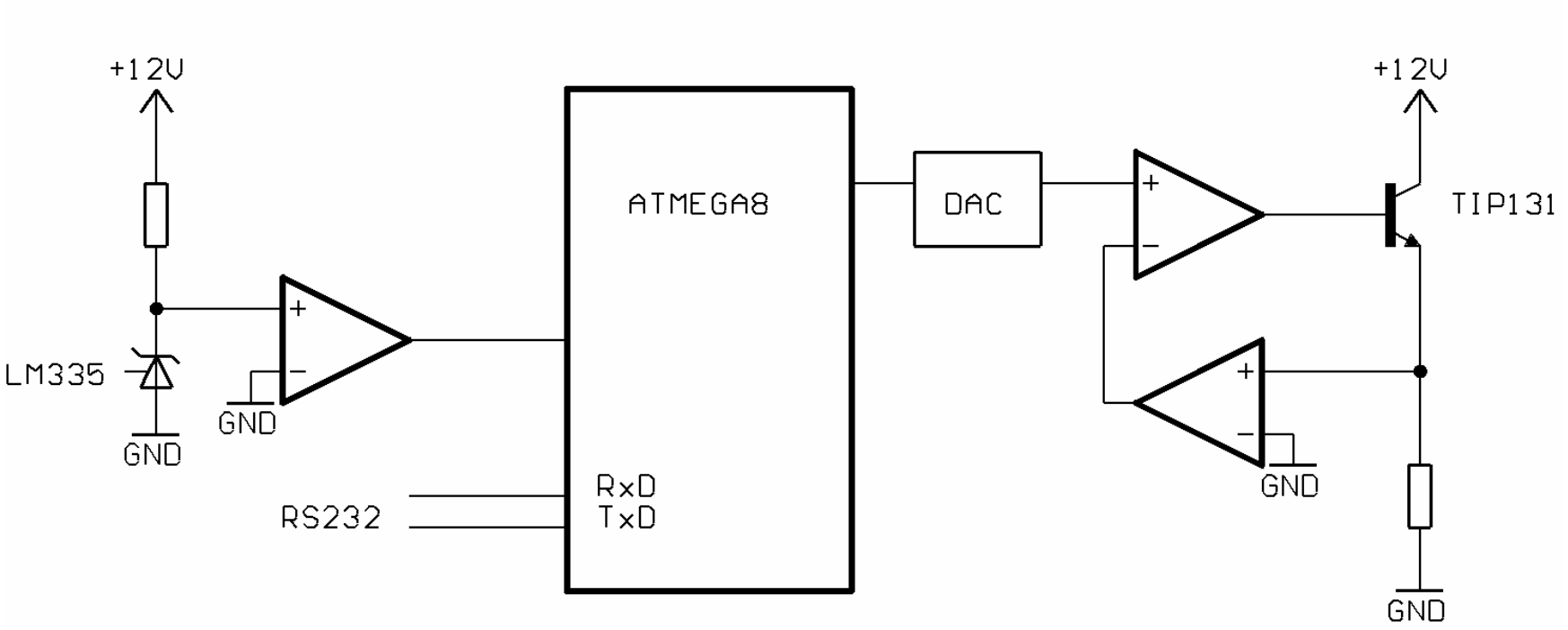

**Figure 5.** Schematic circuit-diagram of heating regulator.

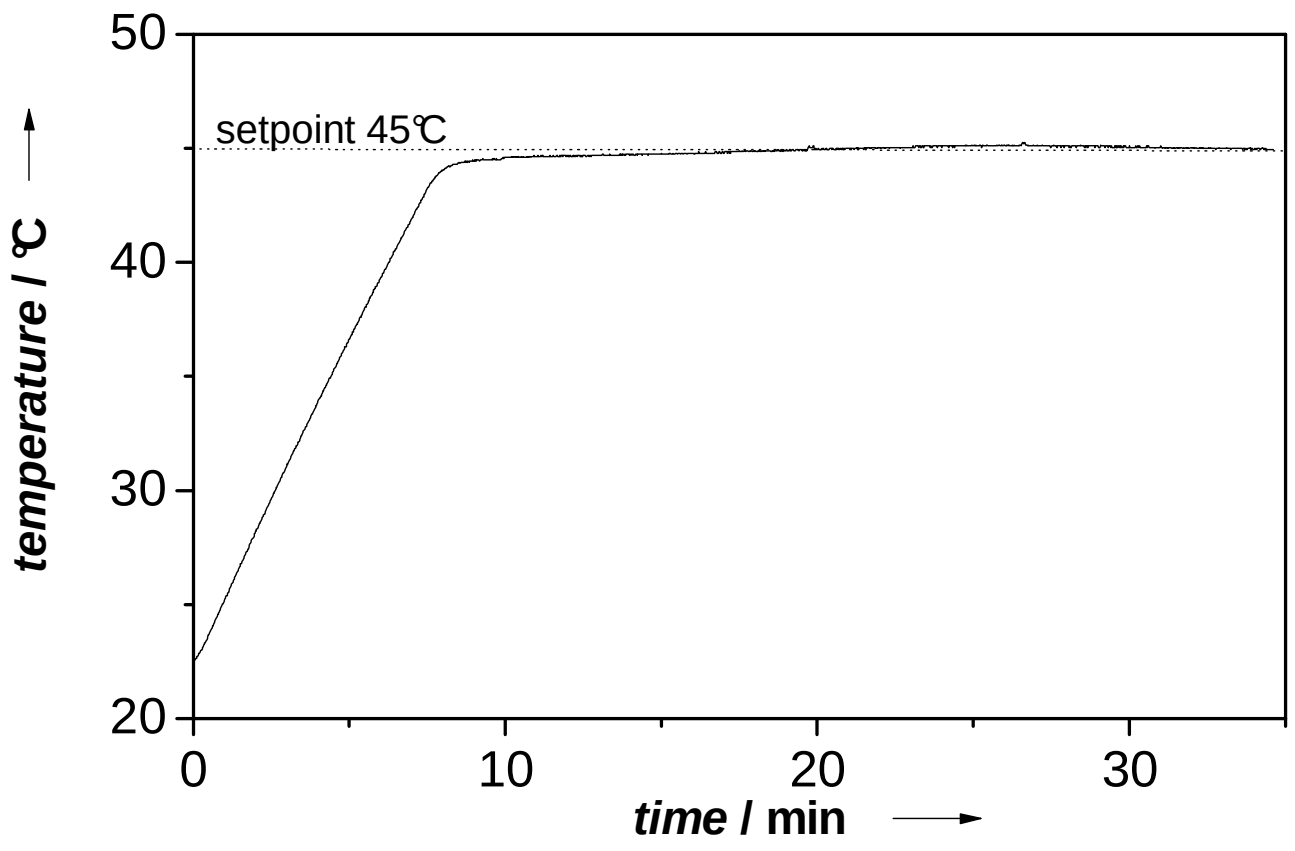

**Figure 6.** Time dependence of temperature during heating of the copper block to a set-point temperature of 45°C.

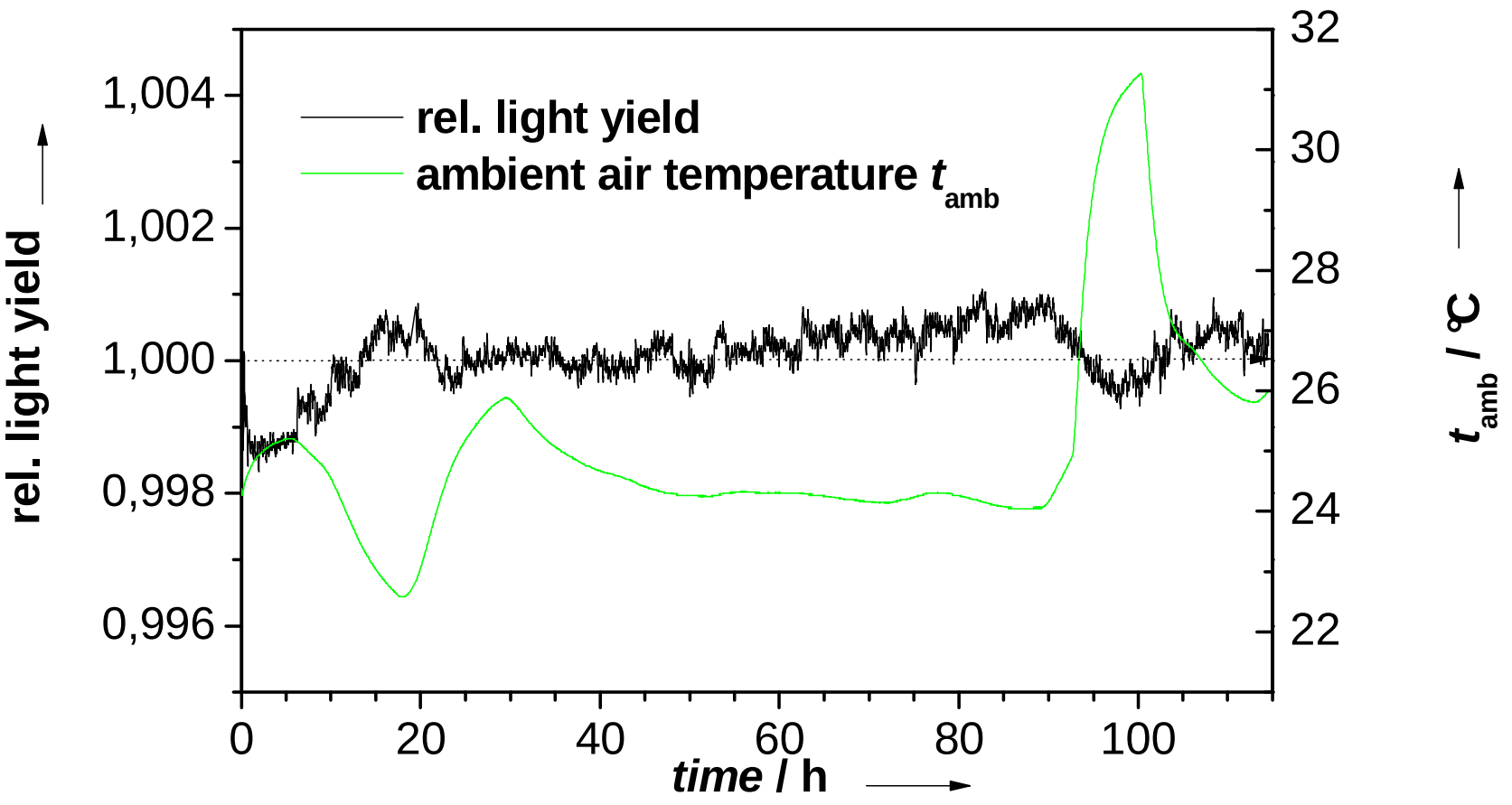

**Figure 7.** The combination of temperature regulation and thermal insulation ensures that ambient temperature fluctuations give rise to barely-discernable variations in LED light-yield

## 2.2. Constant Current Source

Employing an LED as light reference source requires the use of a constant current source, also operated within the thermally-stabilized environment. The schematic drawing of the source as shown in figure 8. The demands on its' temperature-stability are not very stringent. As reference voltage we use an LM 4040 [7] at $U_{ref}$ = 4.096 V, that drives the LED via a voltage-to-current transformer. The current flowing through the LED is thus determined as $U_{ref}$ / R1. The maximal allowable current is limited by the power dissipation of the BSS 670 transistor [10] (~250 mW) and amounts to ~50 mA.

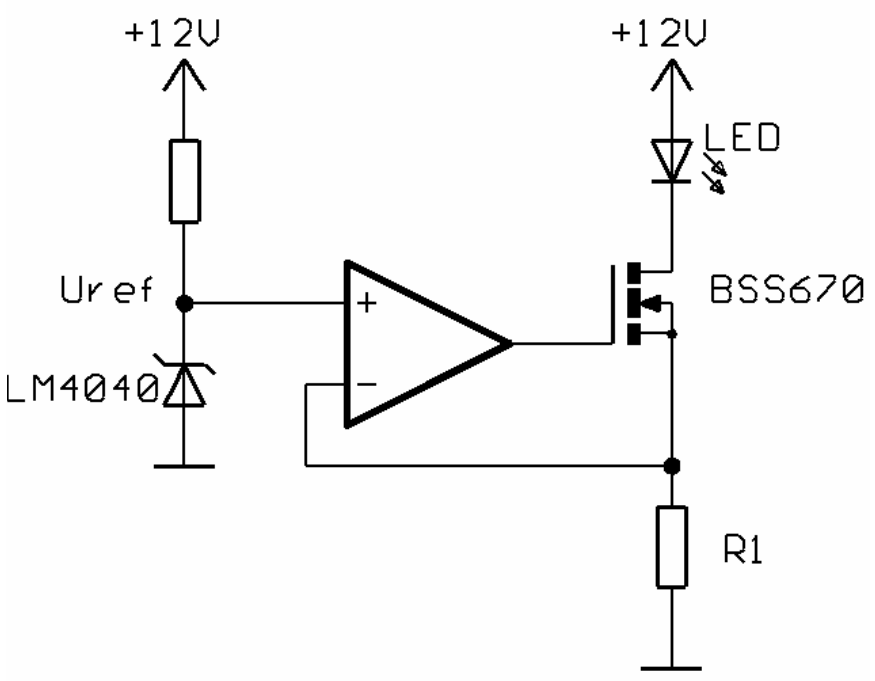

**Figure 8.** Circuit diagram of the constant current source

## 2.3. LED-Monitoring System

Monitoring the temperature regulation and selecting suitable LEDs call for an appropriate measuring system. To this end, we have used a Hamamatsu S1337 photo-diode [11] and attached it to a copper block that, similar to the one in the light source, is maintained at a constant temperature. This measuring system is also thermally shielded inside a custom-made box. The housing of the light-source and photo-diode incorporates locating pins and boreholes, in order to define a precise and easily reproducible measurement geometry (figure 9).

Subsequent measurements have shown that the thermal sensitivity of the photo-diode is far lower than that of the LED. Specifically, temperature jumps from 40 °C to 45 °C and from 25 °C to 40 °C induce variations in the diode sensitivity of $7 \cdot 10^{-4}$ and $1.9 \cdot 10^{-3}$ respectively. These values imply a temperature sensitivity of $\sim 1.4 \cdot 10^{-4}$ / °C. For the present application, it is thus fully sufficient to maintain the temperature of the reference photo-diode stable to within a few Kelvin.

The photo-current is measured by means of a current-integrator, that converts a current to digital pulses at a frequency proportional to the current or, in other words, produces a pulse for a defined and pre-selectable electric charge (typically of the order of 1 pulse / nC). These pulses are counted for one minute and consecutively stored using a data-logger [12].

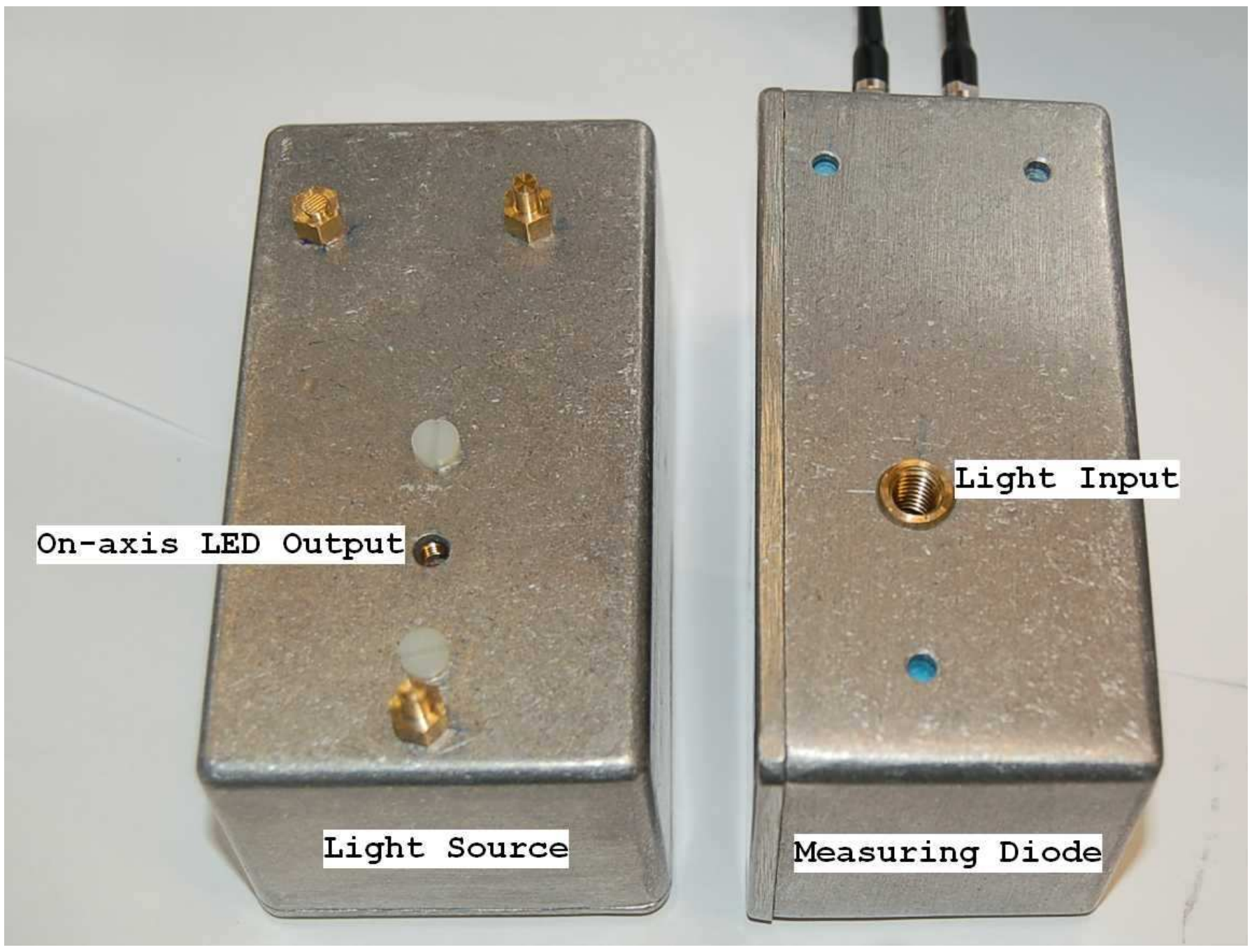

**Figure 9.** Photography of LED source (left) and monitoring system (right). Locating pins (yellow brass on left box) and boreholes (in blue on the right module) ensure reproducible positions of light-source and monitoring photo-diode

## 3. Measurements and results

### 3.1. Ageing of LEDs

It is known that LEDs undergo an ageing process. Moreover, as we experienced during our experiments, these ageing effects cannot be compensated by employing a monitor-diode at an off-axis angle, because the ageing affects not only the total light yield but also the LED's directional emission characteristics. Figure 10 shows the light-yield vs. time curves of a blue Nichia NSPB510S [13] (taken at 13 mA operating current), that had previously been pre-aged for 240 h at 30 mA current. The monitor-diode views the LED at 90°, whereas the usable signal is obtained on the LED's main emission axis in the forward direction.

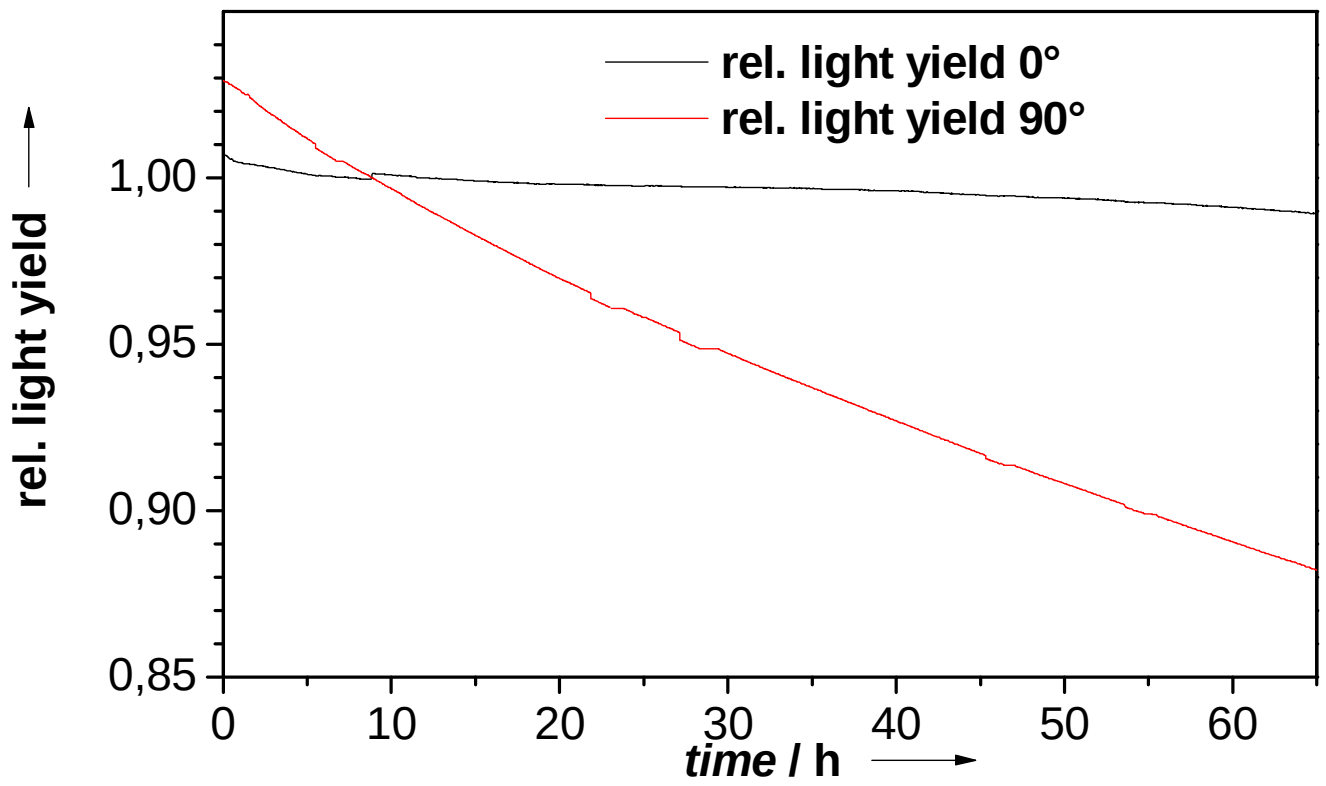

**Figure 10.** Ageing-curves measured at 90° and 0° of the LED's main emission axis

The ageing-effect can be represented, to first approximation, as a time dependent light yield which exponentially approaches a stable end value [14]. Therefore, it should be possible, via a pre-ageing procedure, to operate the LED in a region where the diode light-yield varies only slowly with time. However, when pre-ageing is implemented at the maximal permissible current, the diode subsequently exhibits a "recovery" effect when operated at a lower current. Figure 11 shows the time-behaviour of a Nichia NSPB510S diode, that had been pre-aged for

~600 h at 30 mA current, and was then operated at 10 mA. Here, the duration of the "recovery" process for the 0° emission was ~20 h.

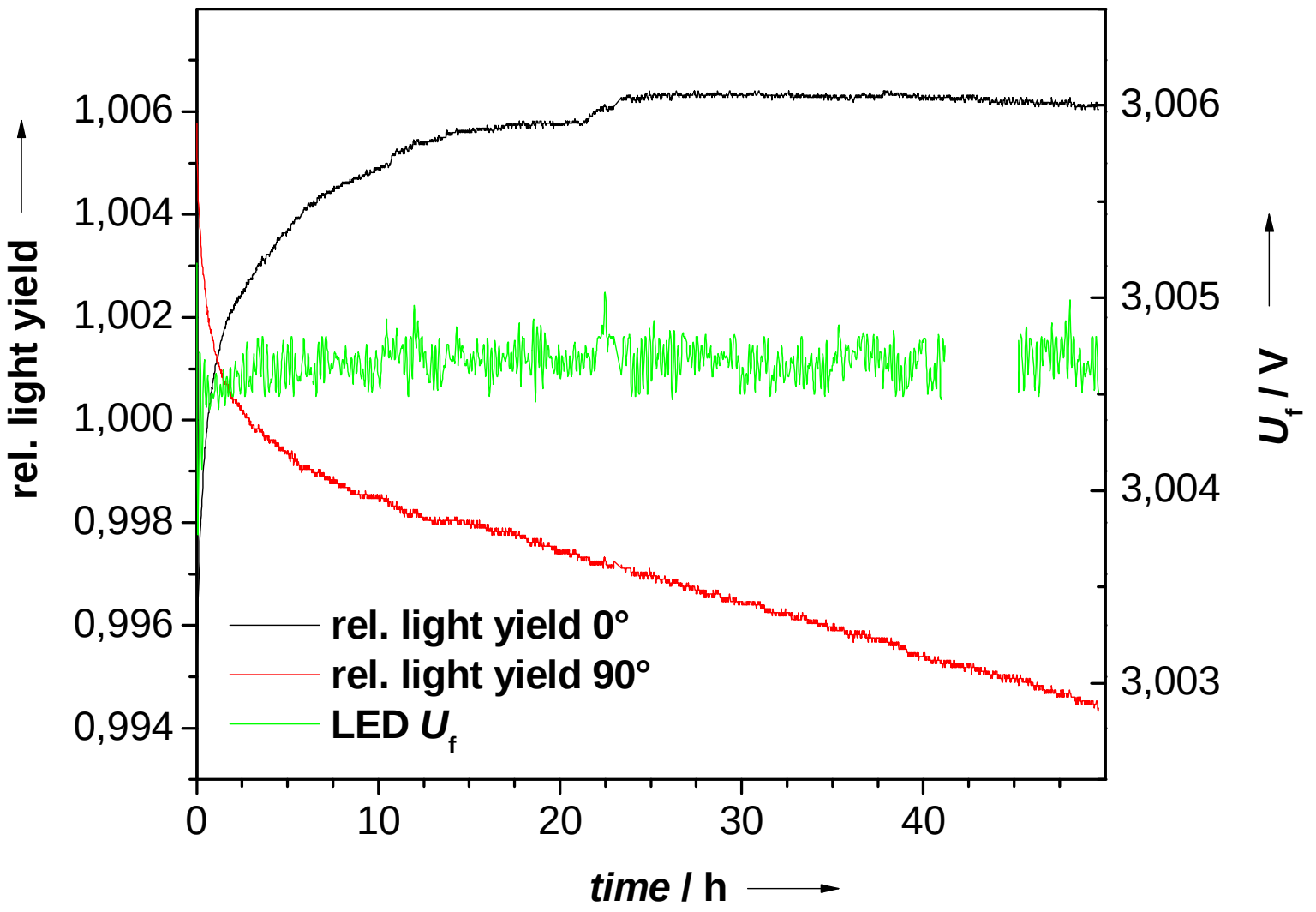

**Figure 11.** For this sample, the "recovery" process following ageing at high current is completed within ~20 h under the given conditions.

### 3.2. Behaviour of different LEDs under same conditions of operation

We compared different LEDs from the same batch in order to understand whether the aging features are reproducible and may be predictable by understanding the underlying mechanism. To this end, several samples of the blue Osram LB543C diode were investigated. The first of these LEDs was pre-aged for 300 h at 30 mA, followed by a monitoring phase of 120 h duration at a current of 12 mA. After a break, the monitoring was continued for another 120 h under the same conditions Figure 12a & b show the results for the first (a) and second (b) monitoring phase. After the break, the 0° light output (black curve) is very stable; in fact, for this diode no further change in the 0° light yield could be observed.

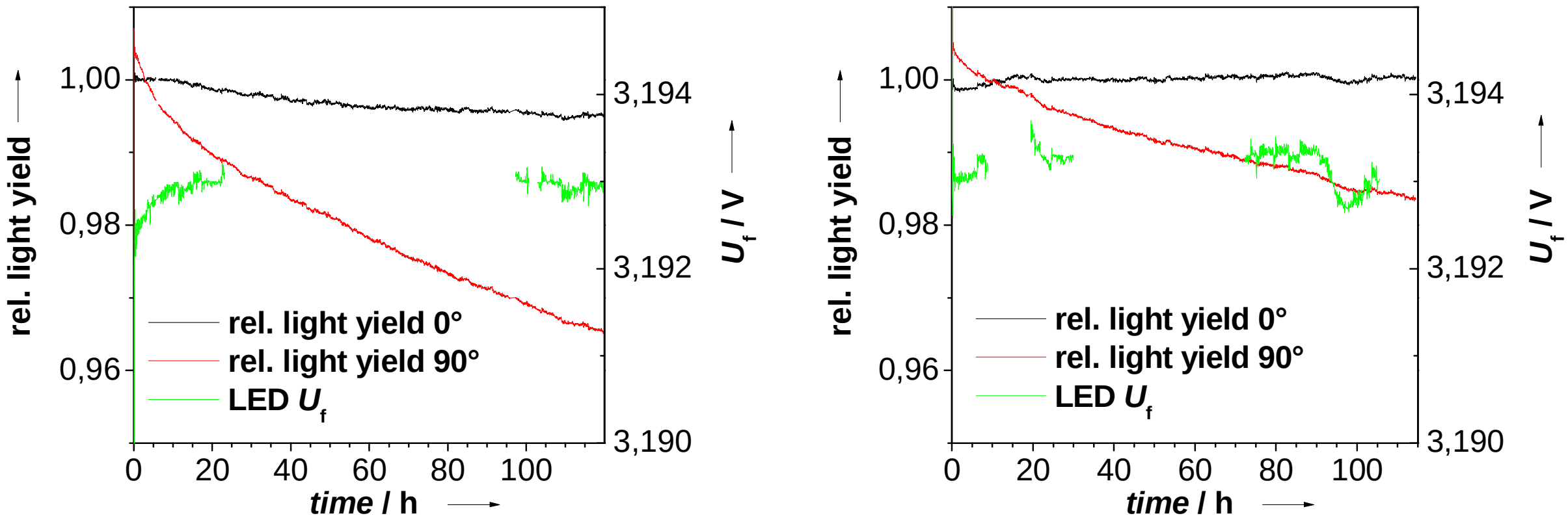

**Figure 12.** 0° and 90° light yield and diode voltage of an LE D, pre-aged for 300 h at 30 mA;
a) monitored for 120 h at 12 mA,
b) after a break and continued monitoring for another 120 h at 12 mA.

The second sample was not pre-aged but continuously operated at 20 mA (figure 13a). After 300 h, its forward-drop-voltage was relatively stable, and yet its forward light-yield continued to increase strongly. Following subsequent ageing for 1500 h at 30 mA, it was operated at 10mA (figure 13b). After about 100 h the recovery process was complete and the light-yield at the external sensor became quite stable. The unexplained jumps indicate a possible defect.

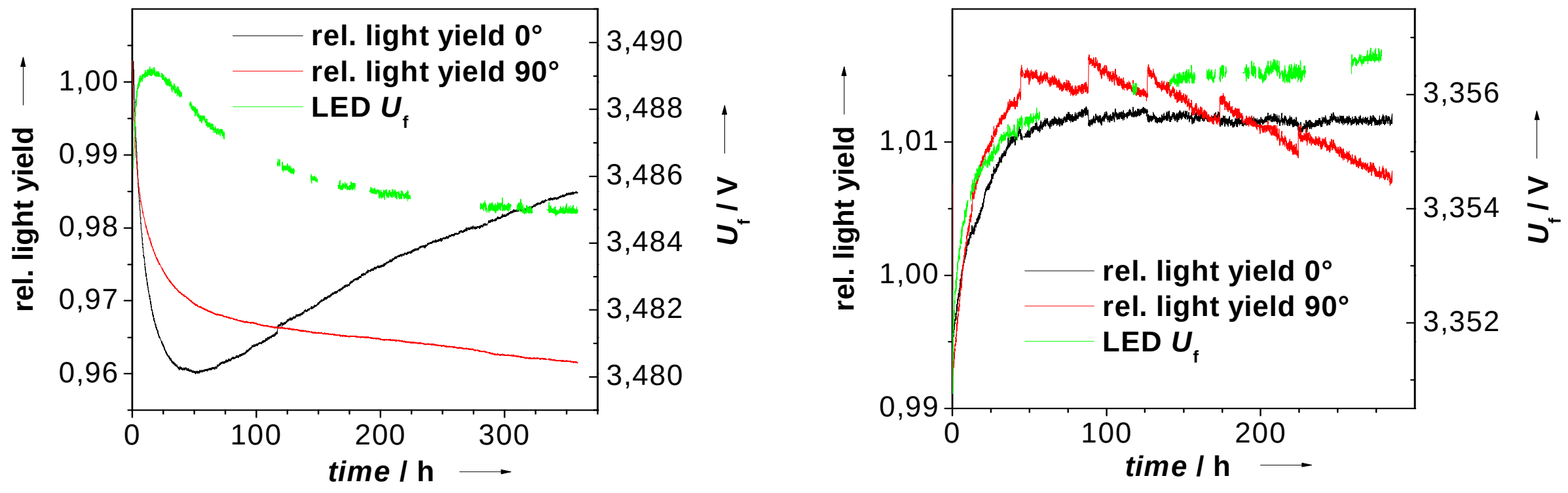

**Figure 13.** Time dependence of forward and 90° light yield and diode voltage of
a) a new LED monitored for 340 h when operated at 20 mA,
b) after aging for 1500h at 30 mA followed by a monitoring phase of 300 h at 10 mA.

A third new and not previously aged LED sample of the same batch was operated at 20 mA for initially 200 h. Figure 14a shows the behaviour during the following 200 h at the same current. Even towards the end of this 2$^{nd}$ aging process the changes in light-yield were too large to be acceptable in our application. Figure 14b shows the characteristics of the same LED after 800 h at 20 mA. While the diode voltage stabilized the light-yields in both directions are still increasing – so this sample also cannot be used in our light source.

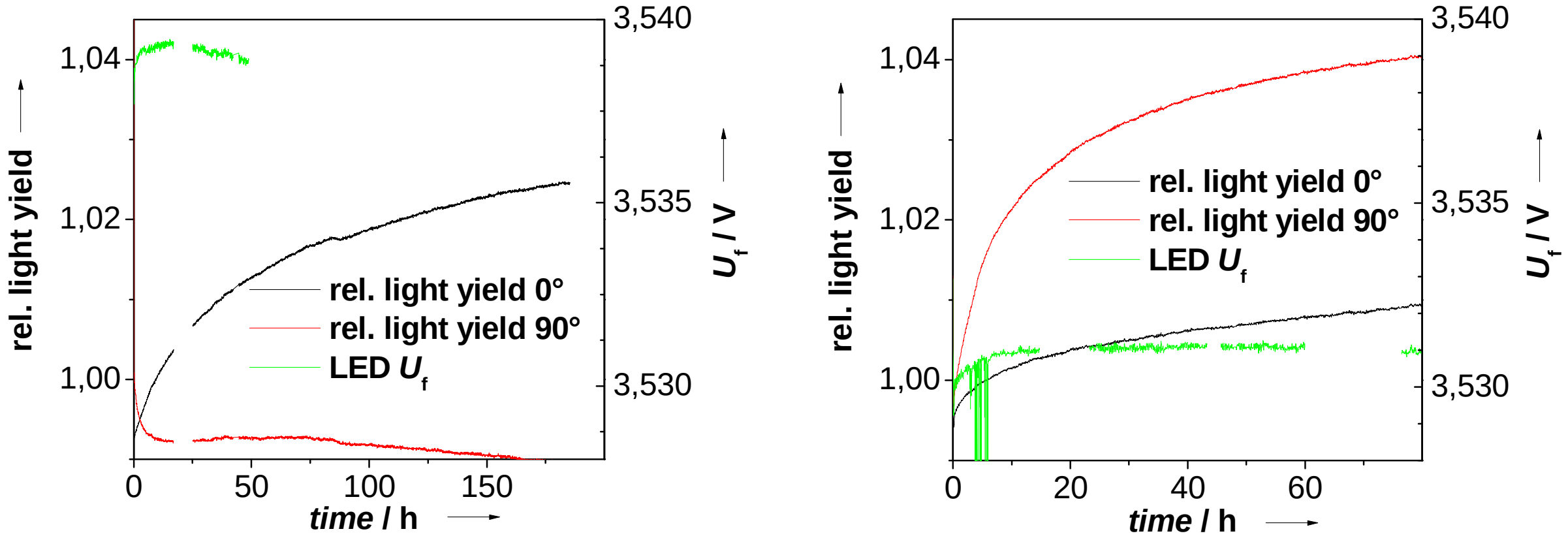

**Figure 14.** Time dependence of forward and 90° light yield and diode voltage of
a) a new LED monitored for 200 h after pre-aging for 200 h,
b) after aging for 800h followed by a monitoring phase of 80 h.
The diode current during aging and monitoring processes was always 20 mA.

In a 4th run another LED from the batch was pre-aged for 400 h and operated at 20 mA (see Figure 15a). The LED forward-drop-voltage indicates that the recovery process was not yet complete. After reduction of the current to 10 mA and a further recovery phase, reasonably good stability could be observed for the forward light yield.

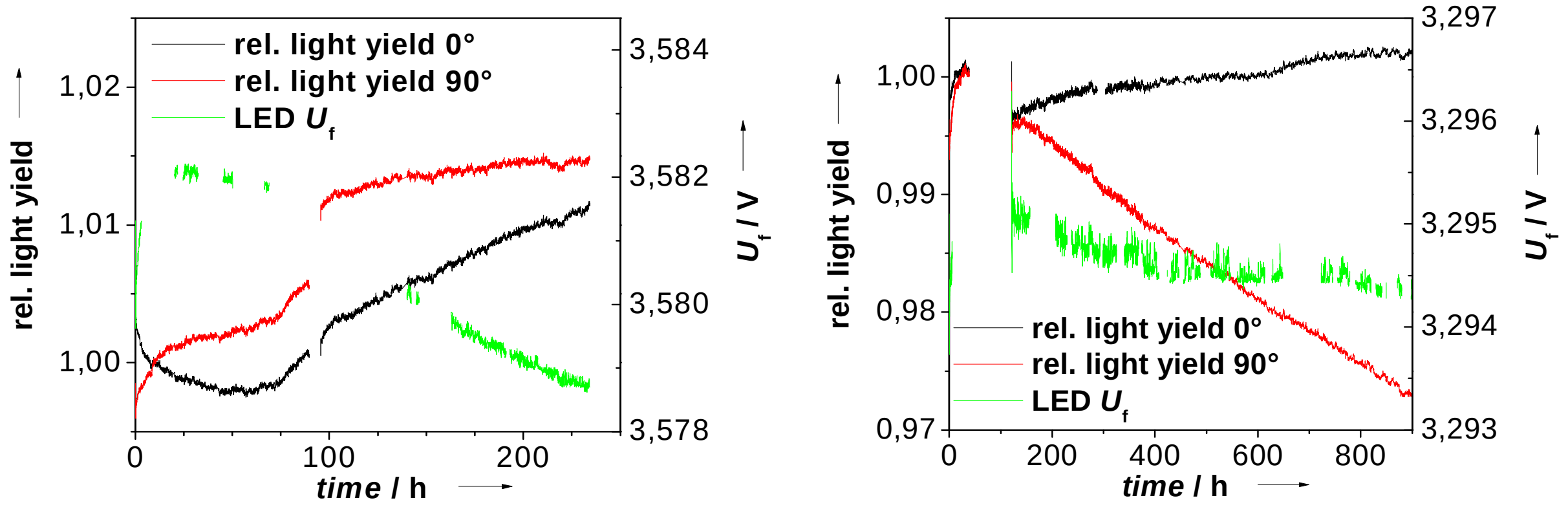

**Figure 15.** Time dependence of forward and 90° light yield and diode voltage of
a) a new LED monitored for 240 h after pre-aging for 400 h at 20 mA;
b) after aging for 600h followed by a monitoring phase of ca 800 h at 10 mA.

## 4. Conclusion

The characteristics of the 4 LED samples in chapter 3.2, all of the same type and taken from the same batch, are just a few examples out of many we have characterized during this investigation. They show that each individual LED behaves differently and that there is no common predictable behaviour of these LEDs. For our application, where we need a LED with long-term stability in its light output, it is required to age several LEDs at close to the maximum current and then select the suitable ones by individual testing and long-term (several 10 to 100 h) monitoring of the light emission characteristics. The criterion for suitability is that changes in the forward light-yield are small and predictable, i.e., the diode exhibits no sudden changes such as those observed in figure 13b.

Since this process is very time consuming and thus costly, one has to look for possibilities of automating it. At the optics department at PTB a system was developed for automatic ageing and monitoring of up to 100 LEDs simultaneously.

Even for a "good" LED, it is strongly recommended to monitor the light yield periodically and verify the predicted behaviour experimentally. For our application in TRION we have adopted a procedure that measures the light yield before and after an experimental run with TRION (typically once a week), to verify reliability and stability of its performance.